\def\half{\frac{1}{2}}
\def\({\left (}
\def\){\right)}
\def\[{\left [}
\def\]{\right]}
\def\<{\left <}
\def\>{\right>}
\documentclass[a4paper,12pt]{article}
\usepackage{amssymb, amsmath}
\usepackage{epsfig}
\makeatletter
\renewcommand{\section}{{\setcounter{equation}{0}}\@startsection%
{section}%
{1}%
{0mm}%
{-\baselineskip}%
{0.5\baselineskip}%
{\normalfont\normalsize\bfseries}%
} \makeatother

\newcommand{\ben}{\begin{enumerate}}
\newcommand{\een}{\end{enumerate}}
\newcommand{\be}{\begin{equation}}
\newcommand{\ee}{\end{equation}}
\newcommand{\bea}{\begin{eqnarray}}
\newcommand{\eea}{\end{eqnarray}}
\newcommand{\beas}{\begin{eqnarray*}}
\newcommand{\eeas}{\end{eqnarray*}}
\newcommand{\begth}{\begin{theorem}}
\newcommand{\enth}{\end{theorem}}
\newcommand{\blem}{\begin{lemma}}
\newcommand{\elem}{\end{lemma}}
\newcommand{\non}{\nonumber}
\newcommand{\nl}{\newline}
\newtheorem{theorem}{Theorem}[section]
\newtheorem{lemma}{Lemma}[section]

\def\RR{\mathbb{R}}

\def\CC{\mathbb{C}}

\def\ZZ{\mathbb{Z}}

\def\la{\langle}
\def\ra{\rangle}
\def\veps{\varepsilon}
\begin{document}
\markboth{A Dicke Type Model for Equilibrium BEC Superradiance} {A
Dicke Type Model for Equilibrium BEC Superradiance}
\phantom{.} \textbf{JSP 104-136} \vskip2cm
\begin{center}
{\bf A Dicke Type Model for Equilibrium BEC Superradiance} \vskip
0.5cm {\bf Joseph V. Pul\'e}  \footnote{{\it
Research Associate, School of Theoretical Physics, Dublin
Institute for Advanced Studies.}}
\linebreak
Department of Mathematical Physics
\linebreak
University College Dublin\\Belfield, Dublin 4, Ireland \linebreak Email:
Joe.Pule@ucd.ie \vskip 0.3cm
{\bf Andr\'{e} F. Verbeure}
\linebreak
Instituut voor Theoretische Fysika,
\linebreak
Katholieke Universiteit Leuven, Celestijnenlaan 200D,\\ 3001 Leuven, Belgium
\linebreak Email:
andre.verbeure@fys.kuleuven.ac.be \vskip 0.3cm
and
\vskip 0.5cm {\bf Valentin A. Zagrebnov} \linebreak Universit\'e de la M\'editerran\'ee and
Centre de Physique Th\'eorique \linebreak CNRS-Luminy-Case 907
\linebreak 13288 Marseille, Cedex 09, France \linebreak Email:
zagrebnov@cpt.univ-mrs.fr
\end{center}
\vskip 1cm
\begin{abstract}
\vskip -0.4truecm
\noindent
We study the effect of electromagnetic radiation on the condensate  of a Bose gas. In an earlier paper
we considered the problem for two simple models showing the cooperative effect between
Bose-Einstein condensation
and superradiance. In this paper we formalise the model suggested by Ketterle et al in which the Bose
condensate particles have a
two level structure. We present a soluble microscopic Dicke type model describing a
thermodynamically stable system.
We find the equilibrium states of the system and compute the thermodynamic functions giving explicit
{formul\ae} expressing the cooperative effect between
Bose-Einstein condensation and superradiance.
\nl {\bf  AMS:} 82B10 , 82B23,  81V80
\end{abstract}

\newpage\setcounter{page}{1}
\section{Introduction}
The present paper is motivated by the recent experiments exhibiting a special coherent
interaction between matter and light,
which has been nicknamed \lq\lq four-wave mixing'' \cite{K}.
In these experiments boson atoms with an internal structure, condensed in a trap, are irradiated
with light produced by an external laser beam.
The structure of the atoms is usually represented by considering them as having two levels
\cite{KI-1,KI-2}.
A system of two level-atoms interacting with light is very reminiscent of the Dicke model \cite{D}.
Moreover an important feature of this model,
namely superradiance has been observed in these experiments, where it is found that there
is an enhancement of both Bose-Einstein Condensation (BEC) and light radiation (superradiance) due to
the interaction.
\par
Recently various models \cite{K} - \cite{KI-2} for this
BEC-superradiance coupling were constructed and discussed in order
to describe both equilibrium and non-equilibrium superradiance by
condensed atoms. It is interesting to note that as early as 1978
Girardeau \cite{Gi} had already anticipated this phenomenon in the
context of superfluid helium and had discussed the possible impact
of the equilibrium superradiance on the thermodynamic properties of
the latter. In our recent letter \cite{PVZ} we have considered two
simple systems by which we modelled the coherent behaviour of the
BEC atoms irradiated by a laser beam, showing rigorously that a
weakened form of the \lq\lq four-wave mixing'' interaction enhances
the superradiance and BEC as proposed by Ketterle et al \cite{K} -
\cite{KI-2}.
\par
The aim of the present paper is to consider a model which takes
explicitly into account the \textit{internal structure} of the boson
atoms. In fact we assume that our bosons have an internal two-level
structure of the type described by $SU(2)$-spin symmetry. Therefore
the one-particle wave functions are of the form $\psi\otimes s$
where $\psi\in L^2(\RR^\nu)$ describing the spacial localization and
$s\in \CC^2$ describing the internal (spin) state. Only the
condensate particles, i.e., the particles in the ground state are
supposed to interact with the external field, and therefore only the
ground state boson particles are given a different ground state
energy parameterized by a separation level parameter $\veps$. If
$\veps$ is put equal to zero, it is as if we have just two different
types of boson particles.  The interaction turns out to be a second
quantized version of the well known Dicke maser model. In our model
we suppose that the \textit{recoil} of the particles is negligible.
The model is in fact a realization of the physical mechanism
explained in \cite{KI-1}. For our model we study the equilibrium
states in the infinite volume limit (thermodynamic limit) and
compute the corresponding thermodynamic functions. We examine the
presence of cooperation between the BEC condensate and superradiance
as a function of the separation level parameter $\veps$. The
existence of this phenomenon confirms the results obtained in
\cite{PVZ} for a simpler model. It can be seen explicitly from the
expressions for the occupation densities for the bosons and photons.
Our results predict that with conventional BEC one obtains the same
phenomenon of BEC-superradiance cooperation as is observed for trap
experiments.

We note that experimentally one can observe the photon recoil effect
which, on light atoms, can be non-negligible \cite{K}. However in
the present paper we consider the case when the photon momentum is
very small so that the recoil effect can still be neglected. In a
later publication we shall study another model in which the
influence of recoil is included.
\section{The model and its equilibrium states}
We consider a system of \textit{two types} of bosons of mass $m$
enclosed in a cubic box $\Lambda$ in $\nu$ dimensions
($\Lambda\subset \RR^\nu$) with volume V, centered at the origin.
\par
As usual let $\Lambda^*=\{2\pi k /V^{1/\nu}| k\in\ZZ^\nu\}$ be the
dual space of $\Lambda$ used to formulate the model with periodic
boundary conditions. For $k\in \Lambda^*$, $\sigma=\pm$,
$a^*_{k,\sigma}$ and $a_{k,\sigma}$ are the usual boson creation
and annihilation operators of the two types of bosons satisfying
the commutation relations: \be
[a_{k,\sigma},a^*_{k',\sigma'}]=\delta_{k,k'}\delta_{\sigma,\sigma'}.
\ee The kinetic energy of the system is given by
\begin{equation}\label{kinetic}
T_\Lambda=\sum_{\sigma=\pm}\ \sum_{k\in \Lambda^*, \, k\neq
0}\epsilon(k)a^*_{k,\sigma}a_{k,\sigma}+\veps(a^*_{0,+}a_{0,+}-a^*_{0,-}a_{0,-})\,,
\end{equation}
where $\veps\geq 0$ and $\epsilon(k)=\|k\|^2/2m$. Note that the two
$k=0$ mode bosons (the ground state for non-interacting bosons) have
a supplementary internal energy, a spin-state energy, making the
internal structure of the bosons explicit. On the other hand the
excited bosons $k\neq 0$ are not distinguished by their internal
energy, but it is straightforward to make them also distinguished.
The reader will able to see that that our arguments cover also the
situation, when the single particle boson spectrum is presented by
\textit{two} branches: $\epsilon_{\sigma}(k):= \epsilon(k) +
\sigma\,\varepsilon $ for two internal states os bosons.

We represent the external one mode laser field by a single mode
boson field with creation and annihilation operators $b$, $b^*$
satisfying  $[b,b^*]=1$. As we indicated in the introduction here we
consider the case, when the photon momentum is very small so that
the recoil effect is \textit{negligible}. In this approximation we
can take $k=0$ and then
\begin{equation}\label{k=0}
b = \frac{1}{\sqrt{V}}\int_{\Lambda} dx \,b(x) \,,
\end{equation}
where $b(x) ,  x\in \mathbb{R}^\nu$ stands for the local
(annihilation) photon field. As suggested in the introduction we
define our model Hamiltonian as \be H_\Lambda=T_\Lambda+U_\Lambda
\label{Ham} \ee where \be U_\Lambda=\frac{g}{2\sqrt V}
(a^*_{0+}a_{0 -}b +a_{0+}a^*_{0 -}b^*)+\Omega\, b^*b + \frac{
\lambda}{2V} N_\Lambda^2 \label{Int} \ee
and
\[ N_\Lambda=\sum_{k\in \Lambda^*}N_k \,,\,\,
N_k=(N_{k,+}+N_{k,-})\,,\,\,\,
N_{k,\sigma}=a^*_{k,\sigma}a_{k,\sigma}
\]
are respectively total boson number operator, the $k$-boson number
operator  and the boson number operator for momentum $k$ and type
$\sigma$.
\par
Furthermore $\Omega>0$ is the laser frequency and
$g$ is the coupling constant of the interaction between the
bosons and the external field.
Note that without loss of generality we can take $g$ to be positive as we can always
incorporate the argument of $g$ into $b$ by a gauge transformation.
\par
Notice in (\ref{Int}) the presence of the mean-field
repulsive particle interaction with a positive
coupling constant $\lambda>0$. This term is essential in order to
obtain a model describing a thermodynamically stable system, i.e.
ensuring the right thermodynamic behaviour. Indeed one can check
by considering the interaction $U_\Lambda$ in (\ref{Int}), that
\bea U_\Lambda &=&\Omega\, ( b^*+\frac{g}{2\Omega\sqrt V}
a_{0+}a^*_{0 -})(b+\frac{g}{2\Omega\sqrt V} a^*_{0+}a_{0 -})
-\frac{g^2}{4 \Omega V} N_{0-}(N_{0 +}+1)+\frac{ \lambda}{2V} N_\Lambda^2\non\\
&\geq & \frac{ \lambda}{2V} N_\Lambda^2-\frac{g^2}{4 \Omega V}
N_{0-}(N_{0 +}+1). \label{lbound} \eea On the basis of the trivial
inequality $4ab\leq (a+b)^2$, the lower bound of (\ref{lbound}) is
bounded from below, if $\lambda > g^2/8\Omega $, that is if the
stabilizing repulsive interaction coupling constant $\lambda$ is
large with respect to the coupling constant $g$ or if the laser
frequency $\Omega$ is large enough. Therefore we assume that
$\lambda > g^2/8\Omega $ is satisfied for the model (\ref{Ham}). The
reader will see all along in the explicit analysis of the model
below, the importance of this stabilizing condition. We note that,
so far, neither the coherent recoil model, nor the \lq \lq four-wave
mixing'' model nor Girardeau's model are thermodynamically stable,
although Girardeau in \cite{Gi} has stressed the importance of this
stabilization. The models in \cite{PVZ} are stable because of the
linearity of the interaction.
\par
In the present paper we study the equilibrium
states of the model (\ref{Ham}) in the grand-canonical ensemble
and therefore we shall work with the Hamiltonian \be
H_\Lambda(\mu)=H_\Lambda-\mu N_\Lambda \label{Ham-mu} \ee where
$\mu$ is the chemical potential. Specifically our objective is to
identify the infinite volume equilibrium states corresponding to
the Hamiltonian (\ref{Ham-mu}) for a system of three different
types of bosons. One way of achieving this goal is through the
basic variational principle of statistical mechanics. Before
starting to do this we prefer to reformulate the model with the
purpose of showing that our model (\ref{Ham}) is nothing but a
second quantized bosonic form of the Dicke model and hence it
realizes the ideas proposed in \cite{KI-1} and \cite{KI-2}.
\par
We have a system of atoms with internal states
$\sigma=\pm$. Dicke regarded the two-level atom as a spin-1/2
system. This is what we shall also do and therefore we start from
a two-dimensional representation of the Pauli matrices generating
the Lie algebra of $SU(2)$, given by \be \sigma^+ = \(
\begin{matrix}
0 & 1 \\ 0 & 0
\end{matrix}
\),
\ \ \sigma^- = \(
\begin{matrix}0 & 0 \\  1 & 0
\end{matrix}\),
\ \ \sigma^3 = \(
\begin{matrix} 1 & 0 \\ 0 & -1
\end{matrix}\)
\ee and the basis vectors $\{e_+=(1,0),\, e_-=(0,1)\}$ of $\CC^2$
diagonalizing $\sigma^3$.
\par
The one-particle space of bosons is ${\cal H}=L^2(\RR^\nu)\otimes \CC^2$.
Let $f\otimes s$ be an element of ${\cal H}$, then $a^*(f\otimes s)$
is the creation operator of a boson particle with state
vector $f\otimes s$. In particular one can make the following identifications.
\be
a^*_{k,\,\pm}=a^*(f_k\otimes e_\pm)
\ee
where for $k\in \Lambda^*$, $f_k$ is the plane wave function
\be
f_k(x)=\frac{1}{\sqrt V}\,e^{ik\cdot x},\ \ \ x\in \RR^\nu.
\ee
In particular we have
\be
a^*_{0,\,\pm}=a^*(f_0\otimes e_\pm).
\ee
For any $\phi \in {\cal H}$, the creation and annihilation operators $a^* (\phi)$ and $a(\phi)$
are linearly defined on arbitrary
 $n$-particle subspaces of Fock space ${\cal F}({\cal H})$:
\be
a^*(\phi) {\rm {\rm \bf  sym}}(\phi_1 \otimes \phi_2 \otimes
\ldots \otimes \phi_n) = (n + 1)^\half {\rm {\rm \bf  sym}}(\phi
\otimes \phi_1 \otimes \ldots \otimes \phi_n) \ee and \be a(\phi)
{\rm \bf  sym} (\phi_1 \otimes \phi_2 \otimes \ldots \otimes
\phi_n)= n^{- \half} \sum^n_{r=1}\la\phi, \phi_r\ra_{\cal H}
\,\,{\rm {\rm \bf sym}}(\phi_1 \otimes \ldots \otimes {\widehat
\phi_r} \otimes \ldots \otimes \phi_n),
\ee
where ${\rm \bf  sym}$
denotes symmetrization,  $\la \cdot,\cdot \ra_{\cal H}$
is the scalar product in ${\cal H}$,  and $\widehat{\phi_r}$ means that $\phi_r$ is
omitted.
\par
Applying these definitions for $a^\#_{0,\,\pm}$ on the
$n$-particle $k=0$ mode states and using the identity
$\sigma^+ s=\la e_-, s\ra_{\CC^2}\,\, e_+$ we obtain
\bea
&&\hskip -1cm a^*_{0+}a_{0-} \, {\rm {\rm \bf  sym}}
((f_0\otimes s_1)\otimes (f_0\otimes s_2)\otimes\ldots \otimes(f_0\otimes s_n))\non\\
&& \hskip 3cm = \sum^n_{r=1}\sigma^+_r{\rm {\rm \bf
sym}}((f_0\otimes s_1)\otimes (f_0\otimes s_2)\otimes\ldots
\otimes(f_0\otimes s_n)) \eea where \be \sigma^+_r(f_0\otimes
s_1)\otimes (f_0\otimes s_2)\otimes\ldots \otimes(f_0\otimes s_n)=
(f_0\otimes s_1)\otimes (f_0\otimes s_2)\otimes\ldots
\otimes(f_0\otimes \sigma^+s_r)\otimes\ldots \otimes(f_0\otimes
s_n). \ee
 The $k=0$ mode kinetic energy term can be treated similarly. Hence, on the
$n$-particle $k=0$ mode states the sum of the $k=0$ kinetic-energy
term (\ref{kinetic}) and the interaction term with the laser field
(\ref{Int}) takes the form \be \varepsilon \,\, \sum_{i=1}^n
\sigma^3_i +\frac{g}{2\sqrt V}\sum_{i=1}^n (\sigma^+_i
b+\sigma^-_i b^*) \ee which \textit{coincides} with the Dicke
maser model. This proves that the model (\ref{Ham}) (or
(\ref{Ham-mu})) realizes the suggestions of \cite{KI-2}, namely
that it is nothing but a second quantized bosonized form of the
Dicke maser model.
\par
So far we have discussed the structure of
our model. The rest of the section is devoted to the technical
preparation of the basic variational principle of statistical
mechanics applied to our model (\ref{Ham-mu}).
\par
The variational principle states that if ${\cal S}$ is the set of the extremal
translation invariant states and $f$ is the free
energy density defined on ${\cal S}$ by
\be
f(\omega)=\lim_{V\to\infty}\omega(H_\Lambda(\mu)/V)-(1/\beta)S(\omega)
\label{free energy density 1}
\ee
where $S(\omega)$ is the entropy density of the state $\omega$, then a
state $\omega_\beta\in {\cal S}$ satisfying
\be
f(\omega_\beta)=\inf_{\omega\in{\cal S}}f(\omega)
\label{varprinc1}
\ee
is an equilibrium state of (\ref{Ham-mu}) at inverse temperature $\beta$.
\par
The Hamiltonian (\ref{Ham-mu}) is not quadratic in the creation
and annihilation operators, and therefore cannot be diagonalised
by a standard symplectic or Bogoliubov transformation and thus, on
this basis, one is tempted to conclude at first sight that the
model is not soluble. However on closer inspection we find that we
can write (\ref{Int}) in the form
\bea
\frac{U_\Lambda}{V}&=&\frac{g}{2}\left\{
\(\frac{a^*_{0+}}{\sqrt{V}}\)\(\frac{a_{0 -}}{\sqrt{V}}\)\(\frac{b
}{\sqrt{V}}\)
+\(\frac{a_{0+}}{\sqrt{V}}\)\(\frac{a^*_{0 -}}{\sqrt{V}}\)\(\frac{b^*}{\sqrt{V}}\)\right \}\non\\
&&\hskip 2cm +\Omega\,
\(\frac{b^*}{\sqrt{V}}\)\(\frac{b}{\sqrt{V}}\) +\frac{ \lambda}{2}
\(\frac{N_\Lambda}{V}\)^2, \label{Int/V} \eea
so that all the terms
are \textit{space averages}. We have \be
\frac{a_{0\pm}}{\sqrt{V}}=\frac{1}{V}\int_\Lambda dx\, a_{\pm}(x),\
\ \ \frac{a_{0\pm}^*}{\sqrt{V}}=\frac{1}{V}\int_\Lambda dx\,
a_{\pm}^*(x)\ \ \ {\rm and}\ \ \ \frac{N_\Lambda}{V}=\sum_{\sigma
=\pm}\frac{1}{V}\int_\Lambda dx\, a^{*}_{\sigma}(x)\,a(x)_{\sigma}.
\ee and by virtue of (\ref{k=0}), $b^*/\sqrt{V}$ and $b/\sqrt{V}$
are clearly also \textit{space averages}. Without going into all the
mathematical details, the reason why space averages are such a
simplifying feature is that they tend \textit{weakly} to a multiples
of the identity operator \cite{BR}. For example if $\omega$ is a
space homogeneous extremal (mixing) state then for all local
observables, $A$ and $B$ one has \bea \lim_{V\to\infty}\omega\(A\,
\frac{1}{V}\int_\Lambda dx\, a^*(x)\,a(x)B\)&=& \omega\(A\,
B\)\lim_{V\to\infty}\omega\(\frac{1}{V}\int_\Lambda dx\,
a^*_{\sigma}(x)\,a_{\sigma}(x)\)\non\\
&=& \omega\(A\, B\)\omega\(a^*_{\sigma}(0)\,a_{\sigma}(0)\),
\label{state1} \eea so that $N_\Lambda/V$ tends weakly to
$\sum_{\sigma = \pm}\omega\(a^*(0)_{\sigma}\,a_{\sigma}(0)\)$.
Similarly \be
\lim_{V\to\infty}\frac{a_{0\pm}}{\sqrt{V}}=\omega(a_{\pm}(0)),\ \ \
\ {\rm and}\ \ \lim_{V\to\infty}\frac{b}{\sqrt{V}}=\omega(b(0)). \ee
\par
Thus if $\omega\in {\cal S}$, then the contribution  of the term
(\ref{Int}) to the energy density in (\ref{free energy density 1})
yields \bea \lim_{V\to\infty}\frac{\omega(U_\Lambda)}{V}&=&
\frac{g}{2}\left\{ \omega(a^*_{+}(0))\omega(a_{-}(0))\omega(b (0))
+ \omega(a_{+}(0))\omega(a^*_{-}(0))\omega(b^*(0))\right \}\non\\
&&\hskip 2cm +\Omega\, |\omega(b(0))|^2 +\frac{ \lambda}{2}
\(\sum_{\sigma = \pm}\omega\(a^*_{\sigma}(0)\,a_{\sigma}(0)\)\)^2,
\label{state2} \eea The result follows readily from (\ref{state1})
with $A$ and $B$ a multiple of the identity. We can therefore
conclude that in the study of the equilibrium states of (\ref{Ham})
or (\ref{Ham-mu}), we can limit ourselves to searching for solutions
$\omega$ which are product states on the tensor product
\textit{canonical commutation relations} algebra (CCR) of the three
different kinds of particles, namely on \be {\cal A}:={\cal A}_+
\otimes {\cal A}_-\otimes {\cal B} , \label{alg} \ee where ${\cal
A}_\pm$ is the C$^*$ algebra generated by the Weyl operators:
\[W_\pm(f):= \exp \left\{i \,\, \frac{a^*_\pm(f)+
a_\pm(f)}{\sqrt{2}}\right\}\,,\] for all $f\in L^2(\RR^\nu)\cap
L^1(\RR^\nu)$, and ${\cal B}$ by the Weyl operators:
\[W_b(f):= \exp
\left\{i \,\, \frac{b^*(f)+b(f)}{\sqrt{2}}\right\}\,.\] The above
discussion makes it clear that we find the equilibrium states of our
model amongst the states which are determined completely by their
one-point and two-point functions, that is, among the set of
extremal space invariant \textit{quasi-free} states \cite{BR} on the
respective CCR-algebras. This is a consequence of the fact that if
$\omega\in{\cal S}$, the set of states on ${\cal A}$, and ${\tilde
\omega}\in {\cal S}$ is a quasi-free state with the same one-point
and two-point functions as $\omega$, then it follows by Klein's
inequality \cite{Th} that \be S({\tilde \omega})\geq S(\omega). \ee
Therefore since our energy density involves only the one-point and
two-point functions, if ${\cal S}_{QF}$ is the set of quasi-free
state on ${\cal A}$, then \be \inf_{\omega\in{\cal S}}f(\omega)\geq
\inf_{\omega\in{\cal S}_{QF}}f(\omega) \ee and consequently \be
f(\omega_\beta)=\inf_{\omega\in{\cal S}_{QF}}f(\omega). \ee We
denote the set of \textit{quasi-free states} on ${\cal A}_\sigma$ by
$\omega_\sigma$ determined by the constants $\alpha_\sigma$ and the
non-negative operators $A_\sigma$ on $L^2(\RR^\nu)$ and satisfying
\be \omega_\sigma(W_\sigma(f))=\exp\(i \sqrt{2}\,\, \mathfrak{Re}
\(\alpha_\sigma\la 1, f\ra\) - \frac{1}{4}\|f\|^2- \frac{1}{2}\la
f,A_\sigma f\ra\) \label{omega pm} \ee for all $f\in
L^2(\RR^\nu)\cap L^1(\RR^\nu)$, see \cite{BR}.
\par
Note that the states $\omega_\sigma$ are completely determined by
the \textit{one-point function} \be
\omega_\sigma(a_{\sigma}(f))={\bar \alpha}_\sigma\la f, 1\ra \label{a_0pm}
\ee and the \textit{two-point function} \be
\omega_\sigma(a^{*}_{\sigma}(f)a_\sigma(g))=\la g,\,A_{\sigma}
f\ra+|\alpha_\sigma|^2\la 1,f\ra\la g,1\ra \ee for all $f,\,g\in
L^2(\RR^\nu)\cap L^1(\RR^\nu)$.
\par
On ${\cal B}$ we consider the extremal invariant state, which is
determined by one constant $\alpha_b$, see \cite{BR}. \be
\omega_b(W_b(f))=\exp\(i \sqrt{2}\,\, \mathfrak{Re} \(\alpha_b\la 1,
f\ra\) - \frac{1}{4}\|f\|^2\) \label{omega pm} \ee Its one- and
two-point functions are
\begin{equation}\label{a-b}
\omega_b(b(f))={\bar \alpha}_b \left\langle f, 1 \right\rangle
\end{equation}
and
\begin{equation}
\omega_b(b^{*}(f)b(g))=|\alpha_b|^2\left\langle
1,f\right\rangle\left\langle g,1\right\rangle. \label{omega b}
\end{equation}
Note that this one-mode \textit{coherent} state depends only on the
$k=0$ mode. It is possible also to consider a more general
quasi-free states of the form (\ref{omega pm}) to take into account
other photons modes. However since only the $k=0$ mode interacts
with the bosons, this is unnecessary.
\par
Thus the candidates for the
equilibrium states are among the set, ${\cal S}_P$, of products of
quasi-free states, i.e. states of the form \be
\omega=\omega_+\otimes\omega_-\otimes\omega_b. \ee They are
completely parameterized by the set of parameters: $\alpha_\pm,\
\alpha_b\in \CC$ and the integral operators $A_\pm$ on
$L^2(\RR^\nu)$: \be (A_\pm f)(x)=\int_{\RR^\nu}A_\pm(x-y)f(y)d\,^\nu
y, \ \ \ x\in \RR^\nu \ee If ${\hat A}_\pm$ is the Fourier transform
of $A_\pm$, then ${\hat A}_\pm(k)\geq 0$, expressing the positivity
of the states $\omega_\pm$.
\par
The variational principle (\ref{varprinc1}) is now reduced to \be
f(\omega_\beta)=\inf_{\omega\in{\cal S}_P}f(\omega).
\label{varprinc} \ee The entropy density for states in ${\cal S}_P$
is explicitly given by, see \cite{F}: \be
S(\omega)=S(\omega_+)+S(\omega_-), \label{entropy1}\ee where
$S(\omega_b)=0$ because only \textit{one} photon mode is taken into
account. Here \be S(\omega_\pm)=\int_{\RR^\nu} \left\{\(1+{\hat
A}_\pm(k)\)\ln\(1+{\hat A}_\pm(k)\)- {\hat A}_\pm(k)\ln {\hat
A}_\pm(k)\right\}\frac{d\,^\nu k}{(2\pi)^\nu}. \label{entropy2} \ee
A straightforward computation yields: \bea
&&\lim_{V\to\infty}\omega(H_\Lambda(\mu)/V)=-(\mu-\veps)|\alpha_+|^2-(\mu+\veps)|\alpha_-|^2\non\\
&&\hskip 2cm +\int_{\RR^\nu }\(\epsilon (k)-\mu\)\({\hat A}_+(k)+ {\hat A}_-
(k)\)\frac{d\,^\nu k}{(2\pi)^\nu}\non\\
&&\hskip 3cm+\frac{g}{2} \({\bar\alpha}_+\alpha_-{\bar\alpha}_b+\alpha_+
{\bar \alpha}_-\alpha_b\)+\Omega\, |\alpha_b|^2\non\\
&&\hskip 4cm+\frac{ \lambda}{2} \left\{\int_{\RR^\nu }\({\hat
A}_+(k)+ {\hat A}_-(k)\)\frac{d\,^\nu
k}{(2\pi)^\nu}+|\alpha_+|^2+|\alpha_-|^2\right\}^2.
\label{energydens} \eea Note that the pressure $P(\mu)$ of the
system (\ref{Ham-mu}), as a function of the chemical potential
$\mu$, is related to the grand-canonical free-energy
density by \be P(\mu)=-f(\omega_\beta)=-\inf_{\cal S}f(\omega).
\ee
\section{Variational Solutions}
In this section we give a systematic derivation of the equilibrium
states for our model as well as explicit expression for the
corresponding grand-canonical pressure. To this end we solve the
variational principal (\ref{varprinc}), and we start by substituting
(\ref{entropy1})-(\ref{energydens}) into (\ref{varprinc}) to obtain
an expression for the functional $f(\omega)$ in terms of the
variational parameters $\alpha_\pm,\ \alpha_b\in \CC$ and ${\hat
A}_\pm(k)$:
\par
We find that there are \textit{two critical} chemical potentials
$\mu_c^{(1)}(\veps)$ and $\mu_c^{(2)}(\veps)$,
$\mu_c^{(1)}(\veps)\leq\mu_c^{(2)}(\veps)$.
\par
For $\mu<\mu_c^{(1)}(\veps)$,
the two $\sigma = \pm$ Bose gases behave like two mean field Bose
gases with no BEC and they do not interact with the
external $b$-boson laser field, in which there is no condensation either.
\par
For $\mu>\mu_c^{(2)}(\veps)$, there is BEC for the
two $\sigma = \pm$ Bose gases and for the external boson laser field
(superradiance).
\par
When $\mu^{(1)}_c(\veps)<\mu_c^{(2)}(\veps)$, for
$\mu^{(1)}_c(\veps) < \mu < \mu_c^{(2)}(\veps)$ there is BEC only
for the $\sigma = -$ Bose gas.
\par
First we remark that we can take $\alpha_\pm,\ \alpha_b$ real
after a suitable gauge transformation on the boson creation and
annihilation operators $a_{0\pm}$ and $b$, see (\ref{a_0pm}) and
(\ref{a-b}). Note that the squares of these parameters are in fact
the condensate densities of the corresponding boson modes. For
notational convenience we introduce the particle density for an
arbitrary quasi-free state $\omega$ of the form
$\omega_+\otimes\omega_-\otimes\omega_b$, \be \rho:=\int_{\RR^\nu
}\({\hat A}_+(k)+ {\hat A}_-(k)\)\frac{d\,^\nu
k}{(2\pi)^\nu}+|\alpha_+|^2+|\alpha_-|^2=\lim_{V\to\infty}\frac{\omega\(N_\Lambda\)}{V},
\label{rho} \ee that is, $\rho$ is the density of $\sigma = \pm$
particles, excluding the $b$-particles. We get the following
Euler-Lagrange equations for the variational principle
(\ref{varprinc}): \nl Take \be \alpha_+,\ \alpha_-,\ \alpha_b\in
\RR. \ee (i)\ \ Differentiation of $f(\omega)$ with respect to
$\alpha_+$ gives: \be 2\(\lambda \rho+\veps -\mu\)\alpha_+
+g\alpha_-\alpha_b=0, \label{alpha+} \ee (ii)\ \ differentiation
with respect to $\alpha_-$: \be 2\(\lambda \rho-\veps
-\mu\)\alpha_- +g\alpha_+\alpha_b=0, \label{alpha-} \ee (iii)\ \
differentiation with respect to $\alpha_b$, \be 2\Omega\alpha_b
+g\alpha_+\alpha_-=0. \label{alphab} \ee (iv)\ \ and finally
differentiating with respect to ${\hat A}_+$ and ${\hat A}_-$
yields: \be {\hat A}_+(k)={\hat
A}_-(k)=\frac{1}{e^{\beta(\epsilon(k)-\mu+\lambda\rho)}-1}.
\label{A_pm} \ee Note that the last equation implies that $\lambda
\rho -\mu\geq 0 $, since the ${\hat A}_\pm(k)$ are
\textit{positive}. Moreover, the correlation inequality (see e.g.
\cite{FV}) \be \omega\([A^*,[H_\Lambda(\mu),A]]\)\geq 0 \ee for
all observables $A$, applied here with $A=a^*_{0-}$, implies that
$\lambda \rho -\mu\geq \veps \geq 0$. Substituting (\ref{A_pm})
into (\ref{rho}) we get \be \rho=
|\alpha_+|^2+|\alpha_-|^2+2\rho_0(\mu-\lambda\rho) \label{rho2}
\ee where \be \rho_0(\mu):=\int_{\RR^\nu
}\frac{1}{e^{\beta(\epsilon(k)-\mu)}-1}\frac{d\,^\nu
k}{(2\pi)^\nu} \ee is the density of the \textit{free} Bose gas at
chemical potential $\mu$. Recall that $\rho_0(\mu < 0) < \infty$
and that $\rho_0(\mu = 0) < \infty$ for $\nu > 2$.

Solving (\ref{alpha+}), (\ref{alpha-}) and (\ref{alphab}) we have to
distinguish \textit{three} cases:
\par
\begin{figure}[hbt]
\begin{center}
\hskip -0.5cm
\includegraphics[width=15cm]{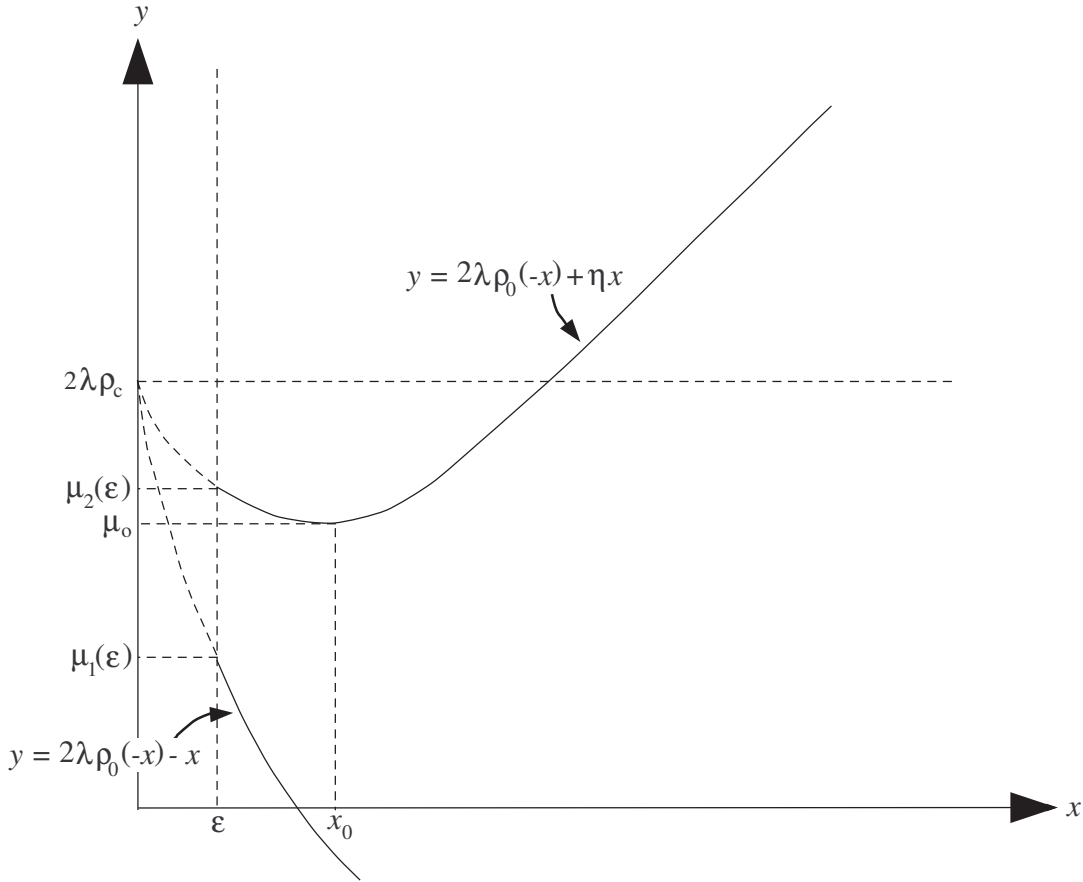}
\end{center}
\vskip- 1cm
\caption{Solution of the density equation} \label{a}
\end{figure}

{\bf Case 1:}\ \ $\alpha_+=\alpha_-=\alpha_b=0$. \nl Substituting
zero for $\alpha_+$ and $\alpha_-$ into (\ref{rho2}) we get the
standard equation for density of the \textit{mean-field interacting}
bosons \be \rho= 2\rho_0(\mu-\lambda\rho), \label{rho Case1} \ee see
e.g. \cite{FV1}. By virtue of the stability condition $\lambda \rho
-\mu\geq \veps \geq 0$ we see that this equation has no solution for
$\mu>\mu_1(\veps):=2\lambda\rho_0(-\veps)-\veps$, while if
$\mu\leq\mu_1(\veps)$ it has a unique solution $\rho=\rho_1(\mu)$.
(See Figure \ref{a}, where $x=\lambda \rho -\mu$ so that $x\geq
\veps$ and (\ref{rho Case1}) becomes $\mu=2\lambda \rho_0(-x)-x$).
 Putting this value of $\rho$ into (\ref{A_pm}), we determine
${\hat A}_\pm$. Substituting these and
$\alpha_+=\alpha_-=\alpha_b=0$ in the expressions (\ref{omega pm}),
(\ref{omega b}) and  (\ref{omega b}), for $\omega_\pm$ and
$\omega_b$ respectively, we find a solution,
$\omega^{(1)}_{\beta,\mu}$, of the Euler-Lagrange equations for the
variational principle (\ref{varprinc}) for $\mu\leq\mu_1(\veps)$.
From (\ref{rho2}) we are able to compute the free energy density for
the state $\omega^{(1)}_{\beta,\,\mu}$: \be
f(\omega^{(1)}_{\beta,\,\mu})=-2p_0(\mu -\lambda \rho_1(\mu))-\half
\lambda \rho_1^2(\mu) \ee where $p_0(\mu)$ is the pressure of the
free Bose gas: \be p_0(\mu):=-\frac{1}{\beta}\int_{\RR^\nu
}\ln\(1-e^{-\beta(\epsilon(k)-\mu)}\)\frac{d\,^\nu k}{(2\pi)^\nu}.
\ee \nl {\bf Case 2:}\ \ $\alpha_+$, $\alpha_-$ and $\alpha_b$ are
non-zero. \nl We obtain from (\ref{alpha+}), (\ref{alpha-}) and
(\ref{alphab}) that \be \alpha_+=\frac{2\sqrt{\Omega\(\lambda \rho
-\veps -\mu\)}}{g},\ \ \ \alpha_-=\frac{2\sqrt{\Omega\(\lambda \rho
+\veps -\mu\)}}{g},\ \ \ \alpha_b=-\frac{2\sqrt{\(\lambda \rho
-\mu\)^2-\veps^2}}{g}. \label{alphas Case2} \ee From these we see
that in this case BEC is indeed present. Again substituting these
values for $\alpha_+$, $\alpha_-$ into (\ref{rho2}) we get \be
\rho=\frac{8\Omega}{g^2}(\lambda \rho-\mu)+2\rho_0(\mu -\lambda
\rho). \label{rho Case2} \ee Note that the first term corresponds to
the condensate density. Let $\eta:=\({8\Omega\lambda}/{g^2}-1\)$.
From the thermodynamic stability condition for (Section 2) we know
that $\eta>0$. Then equation (\ref{rho Case2}) has a \textit{unique}
solution $\rho=\rho_2(\mu)$ for
$\mu>\mu_2(\veps):=2\lambda\rho_0(-\veps)+\eta\veps$. Substituting
this value of $\rho$ into (\ref{alphas Case2}) and (\ref{A_pm}) we
obtain all the parameters $\alpha_+$, $\alpha_-$, $\alpha_b$ and
${\hat A}_\pm$ and consequently we get another solution,
$\omega^{(2)}_{\beta,\,\mu}$, of the Euler-Lagrange equations.
\par
The free energy density for the state $\omega^{(2)}_{\beta,\,\mu}$
can again be computed: \be f(\omega^{(2)}_{\beta,\,\mu})=-2p_0(\mu
-\lambda \rho_2(\mu))-\half \lambda \rho_2^2(\mu) +
\frac{4\Omega}{g^2}(\lambda \rho_2(\mu)-\mu)^2
-\frac{4\Omega\veps^2}{g^2}. \label{free en case2} \ee Denote by
$x_0$ the unique solution of equation $2\lambda\rho'_0(-x)=\eta $
corresponding to the minimum of the function
$2\lambda\rho_0(-x)+\eta x$, and let
$\mu_0=2\lambda\rho_0(-x_0)+\eta x_0$. For $\mu_0<\mu\leq
\mu_2(\veps)$ the equation (\ref{rho Case2}) has two solutions
$\rho=\rho_2(\mu)$ and $\rho={\tilde\rho}_2(\mu)$,
$\rho_2(\mu)>{\tilde\rho}_2(\mu)$. The corresponding states
$\omega^{(2)}_{\beta,\,\mu}$ and ${\tilde
\omega}^{(2)}_{\beta,\,\mu}$ can be found as above. The free energy
density for the state $\omega^{(2)}_{\beta,\,\mu}$ is as in
(\ref{free en case2}) and for ${\tilde \omega}^{(2)}_{\beta,\,\mu}$
it is the same with $\rho_2(\mu)$ replaced by ${\tilde\rho}_2(\mu)$.
\par
{\bf Case 3:}\ \ $\alpha_-\neq 0$ and $\alpha_+=\alpha_b=0$. \nl
From (\ref{alpha+}), (\ref{alpha-}) and (\ref{alphab}) one can see
that this is possible only if \be \rho= \frac{\mu+\veps}{\lambda},
\label{rho Case3} \ee corresponding to the \textit{boundary}
$x=\varepsilon$ of the stability domain, see Figure \ref{a}. The
equation (\ref{rho2}) then requires that $\mu>\mu_1(\veps)$ and
gives \be \alpha_+=\sqrt{\frac{\mu+\veps}{\lambda}-
2\rho_0(-\veps)}. \ee This case corresponds to yet another solution
of the Euler-Lagrange equations, $\omega^{(3)}_{\beta,\,\mu}$, whose
free energy density is given by: \be
f(\omega^{(3)}_{\beta,\,\mu})=-2p_0(-\veps)-\frac{(\mu+\veps)^2}{2\lambda}.
\label{free en case3} \ee We see from above that for certain values
of $\mu$ there are several solutions of the Euler-Lagrange
equations. Since these equations determine only the stationary
points of the free energy functional, if there is more than one such
point, in order to obtain the equilibrium state for a fixed $\mu$ we
have to decide which of the solutions, has the \textit{lowest}
grand-canonical free-energy density.
\par
To proceed with explicit analysis of solutions of the Euler-Lagrange
equations it is easier to work with the variable $x=\lambda
\rho-\mu$ rather than $\rho$. Also in the grand-canonical ensemble
it is more usual to use the pressure instead of the free energy
density. These allow to find the grand-canonical pressure as a
function of its natural variable, the chemical potential. In terms
of $x$ and $\eta$ the equations (\ref{rho Case1}), (\ref{rho
Case2})and (\ref{rho Case3}) become: \be 2\lambda\rho_0(-x)-x=\mu\ \
{\rm for}\ \ \mu\leq \mu_1(\veps), \label{x_1 eps} \ee \be
2\lambda\rho_0(-x)+\eta x=\mu\ \  {\rm for}\ \ \mu\geq \mu_0
\label{x_2 eps} \ee and \be x=\veps\ \ {\rm for}\ \
\mu>\mu_1(\veps). \label{x_3 eps} \ee

\textit{We consider first the case} $\veps=0$. Then
$\mu_1(0)=\mu_2(0)=2\lambda\rho_c$, where $\rho_c:=\rho_0(0)$. So,
in this case the \textit{lower} critical dimensionality the same as
for the free (or mean-field) Bose-gas: $\nu = 2$. The equations
(\ref{x_1 eps}), (\ref{x_2 eps}) and (\ref{x_3 eps}) become : \be
2\lambda\rho_0(-x)-x=\mu\ \ {\rm for}\ \ \mu\leq 2\lambda\rho_c,
\label{x_1} \ee \be 2\lambda\rho_0(-x)+\eta x=\mu\ \ {\rm for}\ \
\mu\geq \mu_0 \label{x_2} \ee and \be x=0\ \  {\rm for}\ \
\mu>2\lambda\rho_c. \label{x_3} \ee In Figure \ref{a} we have drawn
$y=2\lambda\rho_0(-x)-x$ and $y=2\lambda\rho_0(-x)+\eta x$. Recall
that $x_0$ is the unique solution of $2\lambda\rho'_0(-x)=\eta $ and
$\mu_0=2\lambda\rho_0(-x_0)+\eta x_0$. It is easy to see that: \ben
\item For $\mu<\mu_0$, (\ref{x_1}) has a unique solution $x_1(\mu)$
while (\ref{x_2}) does not have a solution. \item In the region
$\mu_0<\mu<2\lambda \rho_c$, (\ref{x_1}) has a unique solution
$x_1(\mu)$ while (\ref{x_2}) has two solutions $x_2(\mu)$ and
${\tilde x}_2(\mu)$, $x_2(\mu)>{\tilde x}_2(\mu)$. \item Finally for
$\mu>2\lambda \rho_c$, (\ref{x_1}) has no solution while (\ref{x_2})
has a unique solutions $x_2(\mu)$ and we also have to consider the
solution (\ref{x_3}), $x=0$. \een
Let
\be
P_1(x,\mu)=2p_0(-x)+\frac{(x+\mu)^2}{2\lambda}.
\ee
and
\be
P_2(x,\mu)=2p_0(-x)+\frac{\{(x+\mu)^2-(\eta+1)x^2\}}{2\lambda},
\ee
Then the situation is as follows:
\ben
\item
For $\mu<\mu_0$, the solution of the variational problem (\ref{varprinc}) is
$\omega^{(1)}_{\beta,\mu}$
and the corresponding pressure
$P(\mu):=-f(\omega^{(1)}_{\beta,\mu})=P_1(x_1(\mu),\mu)$.
\item
For $\mu_0<\mu<2\lambda \rho_c$, the solution of the variational problem is the state
out of $\omega^{(1)}_{\beta,\mu}$,
$\omega^{(2)}_{\beta,\mu}$ and ${\tilde \omega}^{(2)}_{\beta,\mu}$ which minimizes the
free energy density
or equivalently maximizes the pressure.
The pressures for these states are $P_1(x_1(\mu),\mu)$, $P_2(x_2(\mu),\mu)$ and
$P_2({\tilde x}_2(\mu),\mu)$
respectively.
\item
For $\mu>2\lambda \rho_c$, the two candidates for the solution of
the variational problem (\ref{varprinc}) are
$\omega^{(2)}_{\beta,\mu}$ and $\omega^{(3)}_{\beta,\mu}$. The
pressures for these states are $P_2(x_2(\mu),\mu)$ and
$P_3(\mu):=-f(\omega^{(3)}_{\beta,\mu})=2p_0(0)+{\mu^2}/{2\lambda}$.
\een
\begin{figure}[hbt]
\begin{center}
\hskip -0.5cm
\includegraphics[width=16cm]{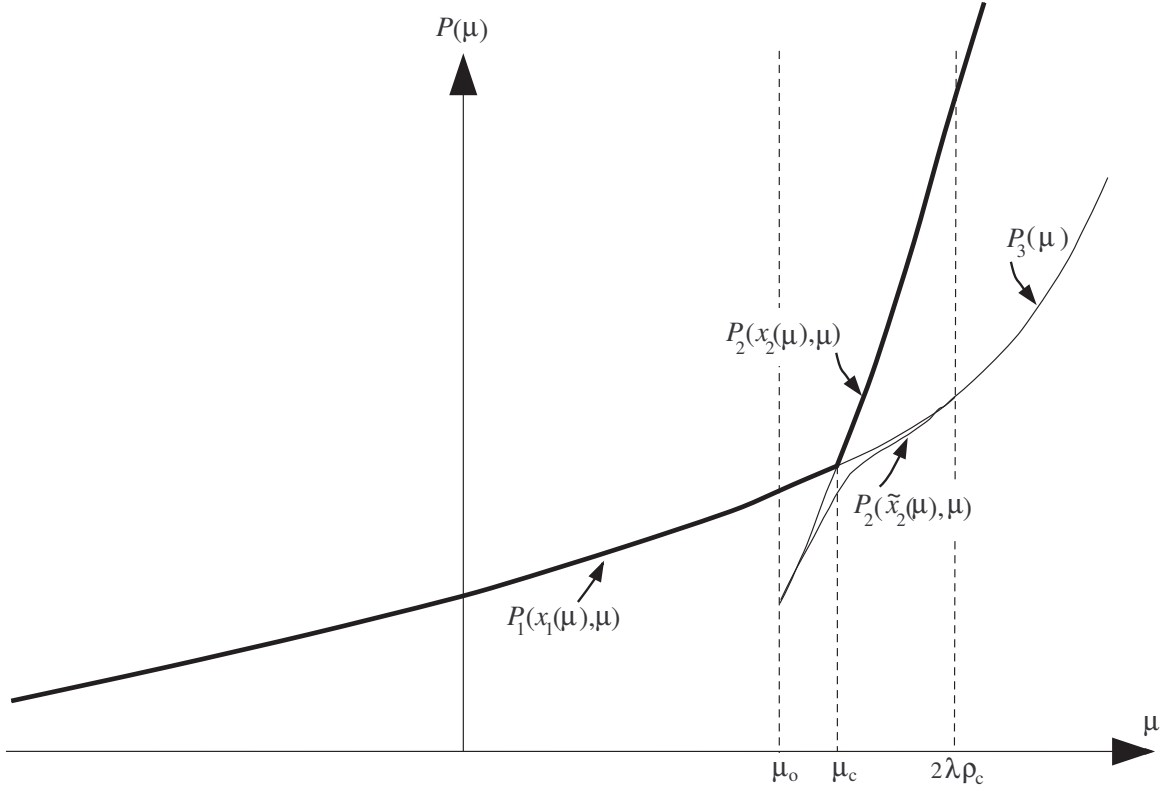}
\end{center}
\vskip- 1cm
\caption{The pressure} \label{b}
\end{figure}
In Figure \ref{b} we have sketched $P_1(x_1(\mu),\mu)$,
$P_2(x_2(\mu),\mu)$, $P_2({\tilde x}_2(\mu),\mu)$ and $P_3(\mu_0)$.
One can check that $P_1(x_1(\mu),\mu)$ and $P_2(x_2(\mu),\mu)$ are
convex in $\mu$. One also has \be
\frac{dP_2(x_2(\mu),\mu)}{d\mu}=\frac{x_2(\mu)+\mu}{\lambda}, \ \ \
\frac{dP_2({\tilde x}_2(\mu),\mu)}{d\mu}=\frac{{\tilde
x}_2(\mu)+\mu}{\lambda} \ee and \be
\frac{dP_1(x_1(\mu),\mu)}{d\mu}=\frac{x_1(\mu)+\mu}{\lambda}. \ee
Therefore since for $\mu_0<\mu<2\lambda \rho_c$, $x_2(\mu)>{\tilde
x}_2(\mu)>x_1(\mu)$, in this interval we have \be
\frac{dP_2(x_2(\mu),\mu)}{d\mu} >\frac{dP_2({\tilde
x}_2(\mu),\mu)}{d\mu}>\frac{dP_1(x_1(\mu),\mu)}{d\mu}.
\label{slopes} \ee As $P_2(x_2(\mu_0),\mu_0)=P_2({\tilde
x}_2(\mu_0),\mu_0)$, it follows from (\ref{slopes}) that
$P_2(x_2(\mu),\mu)>P_2({\tilde x}_2(\mu),\mu)$ for
$\mu_0<\mu<2\lambda \rho_c$. Now \be P_1(x_1(2\lambda
\rho_c),2\lambda \rho_c)=P_2({\tilde x}_2(2\lambda \rho_c),2\lambda
\rho_c)=2p_0(0)+2\lambda \rho_c^2 \label{p rho c} \ee and
consequently $P_2(x_2(2\lambda
\rho_c),2\lambda\rho_c)\!>\!P_1(x_1(2\lambda \rho_c),2\lambda
\rho_c)$. Also if $P_2({\tilde x}_2(\mu_0),\mu_0)$ were greater than
$P_1(x_1(\mu_0),\mu_0)$, then (\ref{slopes}) would imply that
$P_2({\tilde x}_2(2\lambda \rho_c),2\lambda \rho_c)>P_1(x_1(2\lambda
\rho_c),2\lambda \rho_c)$ contradicting (\ref{p rho c}). Thus we
must have \be P_2(x_2(\mu_0),\mu_0)=P_2({\tilde
x}_2(\mu_0),\mu_0)<P_1(x_1(\mu_0),\mu_0). \ee Therefore there exists
a unique $\mu_c$ satisfying $\mu_0<\mu_c <2\lambda \rho_c$ such that
$P_2(x_2(\mu_c),\mu_c)=P_1(x_1(\mu_c),\mu_c)$. \nl Finally we
consider $\mu > 2\lambda \rho_c$. We have
$P_3(2\lambda\rho_c)=P_1(x_2(2\lambda\rho_c),2\lambda\rho_c)<P_1(x_2(2\lambda\rho_c),2\lambda\rho_c)$
and \be
\frac{dP_3(\mu)}{d\mu}=\frac{\mu}{\lambda}<\frac{(x_2(\mu)+\mu)}{\lambda}=
\frac{dP_2(x_2(\mu),\mu)}{d\mu}.
\ee Therefore $P_2(x_2(\mu),\mu)>P_3(\mu)$ for $\mu >2\lambda
\rho_c$.
\par
Summarizing: There exists a unique critical chemical potential
$\mu_c$ such that \ben \item For $\mu<\mu_c$, the solution of the
variational problem (\ref{varprinc}) is
$\omega^{(1)}_{\beta,\mu}$. For $\omega^{(1)}_{\beta,\mu}$,
$\alpha_+=\alpha_-=\alpha_b=0$, i.e. the two $\pm$ Bose gases
behave like two mean field Bose gases with no BEC and do not
interact with the external $b$-bosons which do not condense
either. The corresponding pressure is $P(\mu)=P_1(x_1(\mu),\mu)$.
\item For $\mu>\mu_c$, the solution of the variational problem is
$\omega^{(2)}_{\beta,\mu}$. For this state \be
\alpha_+=\alpha_-=\frac{2\sqrt{\Omega\, x_2(\mu)}}{g}\ \ \ {\rm
and} \ \ \  \alpha_b=-\frac{2x_2(\mu)}{g}, \ee i.e. there is BEC
for the two $\pm$ Bose gases and for the external bosons laser
field (superradiance). Moreover the condensation of the $\pm$
bosons is enhanced by the presence of the laser field
($b$-bosons), known as the equilibrium BEC superradiance
\cite{PVZ}. The pressure for the system is
$P(\mu)=P_2(x_2(\mu),\mu)$. \een

\textit{We now return to the case }$\veps>0$. We have to redefine
$P_2$ and $P_3$ but $P_1$ remains unchanged: \be
P_2(x,\mu)=2p_0(-x)+\frac{\{(x+\mu)^2-(\eta+1)x^2\}}{2\lambda}+\frac{(\eta+1)\veps^2}{2\lambda}
\ee and \be P_3(\mu)=2p_0(-\veps)+\frac{(\mu+\veps)^2}{2\lambda}.
\ee Note that \be
P_1(x_1(\mu_1(\veps)),\mu_1(\veps))=P_3(\mu_1(\veps)) \ee and \bea
P_2({\tilde x}_2(\mu_2(\veps)),\mu_2(\veps))&=& P_3(\mu_2(\veps))\ {\rm for}\ \veps <x_0\non\\
P_2(x_2(\mu_2(\veps)),\mu_2(\veps))&=& P_3(\mu_2(\veps))\ {\rm for}\
\veps >x_0. \eea Also by again considering the derivatives \be
P_2(x_2(\mu),\mu)>P_3(\mu) \ee for $\mu>\mu_2(\veps)$ and as before
$P_2(x_2(\mu),\mu)>P_2({\tilde x}_2(\mu),\mu)$ for
$\mu_0<\mu<2\lambda \rho_0 (-\varepsilon)$ in the region where it
applies.
\par
(a)\ The simplest case to consider is when $\veps >x_0$. In this
case for $\mu<\mu_1(\veps)$ only (\ref{x_1 eps}) has a solution
$x_1(\mu)$, for $\mu_1(\veps)<\mu<\mu_2(\veps)$ only (\ref{x_3
eps}) is satisfied i.e. $x=\veps$ and for $\mu>\mu_2(\veps)$ only
(\ref{x_2 eps}) has a solution $x_2(\mu)$. Thus the states are
$\omega^{(1)}_\beta$, $\omega^{(3)}_\beta$ and
$\omega^{(2)}_\beta$ as $\mu$ increases. This means that as we
increase $\mu$ the system goes from no BEC, to BEC for the $\sigma
= -$ bosons only, to BEC for both species and superradiance.
\par
(b)\ When $\veps <x_0$ we have to consider two cases, $\mu_1(\veps)<\mu_0<\mu_2(\veps)$ and
$\mu_0<\mu_1(\veps)$.
\par
In the first case we can use the same arguments as for $\veps=0$ to
show that $P_3(\mu_0)>P_2(x_2(\mu_0),\mu_0)$ and
$P_3(\mu_2(\veps))<P_2(x_2(\mu_2(\veps)),\mu_2(\veps))$. This
implies that there \textit{exists} $\mu_c(\veps)$ between $\mu_0$
and $\mu_2(\veps)$ such that
$P_3(\mu_c(\veps))=P_2(x_2(\mu_c(\veps)),\mu_c(\veps))$. Thus at
$\mu_c(\veps)$ the state changes from $\omega^{(3)}_\beta$ to
$\omega^{(2)}_\beta$. This means that the situation is the same as
for $\veps >x_0$ except that the changes of state occur at
$\mu_1(\veps)$ and at $\mu_c(\veps)$.
\par
For $\veps <x_0$ and $\mu_0<\mu_1(\veps)$ the same argument applies. However we did not determine on
which side of $\mu_1(\veps)$, the value of $\mu_c(\veps)$
lies. Thus we know that there is no BEC for $\mu<\mu_0$ and there is BEC for both types of bosons for
$\mu>\mu_c(\veps)$, but we do not know if the intermediate
phase with BEC for $-$ bosons only is present.
\section{Conclusion: Equilibrium BEC Superradiance}

The above results may be summarized as follows:
\par
There exist \textit{two critical} chemical potentials
$\mu_c^{(1)}(\veps)$ and $\mu_c^{(2)}(\veps)$,
$\mu_c^{(1)}(\veps)\leq\mu_c^{(2)}(\veps)$.
\par
For $\mu<\mu_c^{(1)}(\veps)$, the solution of the variational problem
(\ref{varprinc}) is $\omega^{(1)}_{\beta,\,\mu}$. For the state
$\omega^{(1)}_{\beta,\,\mu}$,
\be
\alpha_+=\alpha_-=\alpha_b=0,
\ee
i.e.
the two $\sigma = \pm$ Bose gases behave like two mean field Bose
gases with no BEC and they do not interact with the
external $b$-boson laser field, in which there is no condensation either.
\par
For $\mu>\mu_c^{(2)}(\veps)$, the solution of the variational
problem is $\omega^{(2)}_{\beta,\,\mu}$. By virtue of (\ref{alphas
Case2}) for this state we have \be 0 < \alpha_+ \leq \alpha_{-} \, \
\ \ {\rm and} \ \ \ \alpha_b\neq 0, \ee i.e. there is BEC for the
two $\sigma = \pm$ Bose gases and for the external boson laser field
(superradiance). Moreover, for $\varepsilon > 0$ the condensation of
the $\sigma = \pm$ bosons is \textit{enhanced} by the presence of
this laser field: one gets it even for dimensions $\nu =1,2$,
because $\rho_0 (-\varepsilon) <\infty $ for $\nu \geq 1$. We
interpret this quantum state as that of equilibrium BEC
superradiance \cite{PVZ}.
\par
When $\mu^{(1)}_c(\veps)<\mu_c^{(2)}(\veps)$, for $\mu_c^{(1)}(\veps)<\mu<\mu_c^{(2)}(\veps)$,
the solution of the variational problem is
$\omega^{(3)}_{\beta,\,\mu}$. For this state we have
\be
\alpha_-\neq 0\, \ \ \ {\rm and} \ \ \ \alpha_+=\alpha_b= 0,
\ee
i.e. there is BEC only for the $\sigma = -$ Bose gas.
\par
\textbf{(a)}\ The simplest case to consider is when $\veps >x_0$.
In this case $\mu_c^{(1)}(\veps)=\mu_1(\veps)$ and $\mu_c^{(2)}(\veps)=\mu_2(\veps)$.
Thus the states are $\omega^{(1)}_\beta$, $\omega^{(3)}_\beta$ and
$\omega^{(2)}_\beta$ as $\mu$ increases. This means that as we
increase $\mu$ we observe three stages: the system goes
from \textit{no} BEC, to BEC for \textit{only}  the $\sigma = -$
bosons, and then to BEC for both $\sigma = \pm$ boson
species and for the laser field (superradiance).
\par
\textbf{(b)}\ When $0<\veps <x_0$ we have to consider \textit{two}
subcases:  $\mu_1(\veps)<\mu_0<\mu_2(\veps)$ and
$\mu_0<\mu_1(\veps)$.
\par
In the first subcase
$\mu_1(\veps)=\mu^{(1)}_c(\veps)<\mu^{(2)}_c(\veps)<\mu_2(\veps)$.
Otherwise the situation is as in (a).
\par
For $\mu_0<\mu_1(\veps)$,  $\mu_1(\veps)<\mu_c^{(2)}(\veps)<\mu_2(\veps)$ but we did
not determine if $\mu^{(1)}_c(\veps)<\mu_c^{(2)}(\veps)$. Thus we do not know if the
\textit{intermediate phase} with BEC for only $\sigma = -$ bosons
is present.
\par
\textbf{(c)}\ If $\varepsilon = 0$, then $\mu^{(1)}_c(\veps)=\mu_c^{(2)}(\veps)$ and the
\textit{intermediate phase} with BEC for only $\sigma = -$ bosons
is not present.
\par
{\bf Acknowledgements:} Two of the authors (JVP and AFV) wish to
thank the Centre Physique Th\'eorique, CNRS-Luminy, where this work
was initiated, for its inspiring hospitality. They also wish to
thank Bruno Nachtergaele for his kind hospitality at the University
of California, Davis, where this work was finalized. JVP wishes to
thank University College Dublin for the award of a President's
Research Fellowship. We thank the referees for some very useful
remarks and suggestions and Teunis Dorlas for discussing the
variational problem with us.
%

\newpage

\textbf{FIGURE CAPTIONS}

\vskip 0.5cm

Figure 1:  Solution of the density equation

\vskip 0.5cm

Figure 2:  The pressure

\end{document}